\documentclass[aps,prl,twocolumn,superscriptaddress,notitlepage]{revtex4-2}
\usepackage{CJK}
\usepackage{graphicx}
\usepackage{caption}
\usepackage{float}
\usepackage{amsmath}
\usepackage{amssymb}
\usepackage{url}
\usepackage{natbib}
\usepackage{subcaption}
\usepackage{overpic}
\usepackage{tikz}
\usepackage{amsmath}
 
\usepackage[colorlinks,
                   linkcolor=blue,
                   anchorcolor=blue,
                   citecolor=blue,
                   urlcolor=blue]
                  {hyperref}              
\begin{document}
\begin{CJK*}{UTF8}{gbsn}
\title{Ultralow Frequency Magnetic Comb Using Pulse Modulated Spin Maser}
\author{Weiyu Wang}
\thanks{These authors contributed equally to this work and should be regarded as co-first authors.}
\affiliation{National Time Service Center, Chinese Academy of Sciences, Xi'an 710600, China}
\affiliation{Hefei National Laboratory, Hefei 230088, China}
\affiliation{University of Chinese Academy of Sciences, Beijing 100049, China}

\author{Qingchang Li}
\thanks{These authors contributed equally to this work and should be regarded as co-first authors.}
\affiliation{National Time Service Center, Chinese Academy of Sciences, Xi'an 710600, China}
\affiliation{University of Chinese Academy of Sciences, Beijing 100049, China}

\author{Erwei Li}
\affiliation{National Time Service Center, Chinese Academy of Sciences, Xi'an 710600, China}
\affiliation{University of Chinese Academy of Sciences, Beijing 100049, China}

\author{Lan Wu}
\affiliation{National Time Service Center, Chinese Academy of Sciences, Xi'an 710600, China}
\affiliation{University of Chinese Academy of Sciences, Beijing 100049, China}

\author{Qianjin Ma}
\affiliation{National Time Service Center, Chinese Academy of Sciences, Xi'an 710600, China}

\author{Shougang Zhang}
\affiliation{National Time Service Center, Chinese Academy of Sciences, Xi'an 710600, China}
\affiliation{Hefei National Laboratory, Hefei 230088, China}
\affiliation{University of Chinese Academy of Sciences, Beijing 100049, China}
\affiliation{Key Laboratory of Time Reference and Applications, Chinese Academy of Sciences, Xi’an 710600, China}

\author{Guobin Liu}
\email{liuguobin@ntsc.ac.cn}
\affiliation{National Time Service Center, Chinese Academy of Sciences, Xi'an 710600, China}
\affiliation{University of Chinese Academy of Sciences, Beijing 100049, China}
\affiliation{Key Laboratory of Time Reference and Applications, Chinese Academy of Sciences, Xi’an 710600, China}

\date{\today}
\begin{abstract}
Frequency combs are widely used in fundamental physics and practical applications at various areas. Here we report the realization of an ultralow frequency magnetic comb using a pulse modulated spin maser in the Rb-Xe hybrid gaseous spin system. The magnetic frequency comb has a spectral distribution depending mainly on the pulse duration, the strength and phase shift of the feedback field. A simple theoretical model is given in terms of the spin echo chain and agrees well with the experimental results. The magnetic frequency comb works in an ultralow frequency range and reaches a frequency resolution down to tens of nanohertz.

\end{abstract}
\maketitle
\end{CJK*}

\captionsetup[subfloat]{
    font=bf, 
     justification=raggedright, 
    singlelinecheck=false, 
    width=0.8\textwidth, 
    position=top, 
    skip=0pt, 
    margin=5pt
}

\textit{Introduction}--Optical frequency comb, with equally spaced narrow spectral lines along a broad frequency range, opens a new type of precision metrology and finds wide applications from optical atomic clocks to distance measurements and molecular spectroscopy \cite{AO2000LIDAR,CP2019OFC,OPTICA2019clock,NP2022OFCtech,MF2007Nature,MS2008PRL,RS2013Nature}. 
Although the potential applications in sensitive magnetic measurements are not quite obvious, magnetic frequency comb, the magnetic counterpart of optical frequency comb, has inspired many researches recently. Most of the demonstrated magnetic frequency comb relies on the nonlinear magnonics in solid state magnetic materials, namely utilizing the magnon-phonon mutual coupling process with the help of an external pump \cite{PRL2021Skyrmion, APL2022SpinWave,PRL2023MFCresonator, APL2024MFC,NANOLETTERS2024AMFC,NP2024MFCexceptionalpoints,PRApplied2025SpinIce}. 
Solid state magnetic materials are widely used in modern electronic and photonic devices, rendering the magnonic frequency combs good compatibility with matured fabrication techniques. However, the noisy electromagnetic environment inside the solid state materials usually leads to broadened magnetic resonance spectra (with quality factor Q $\sim 10^2$ in \cite{PRL2023MFCresonator}), limiting their potential applications in areas needing ultrahigh precision. Besides, the dissipation of external pump power within the noisy spin system also leads to the low production efficiency of comb lines. 

Compared with spin system in solid states, spins in gaseous atomic media has less electromagnetic noise. Therefore, using vapor atoms, accurate atomic clocks and magnetometers can achieve unprecedent frequency resolution and measurement precision \cite{Ludlow2015RMP,Kominis2003,Budker2007}. Interestingly, the optical frequency comb was originally motivated by the need of counting the cycles of optical atomic clocks. Stimulated by this principle of optical frequency comb, i.e., a pulsed or mode-locked laser, we propose a magnetic frequency comb using the Rb-Xe hybrid spin maser in vapor states. In essence, spin maser is a self-driven spin oscillator by connecting the output of an atomic magnetometer to its input with a feedback circuit. By controlling the strength and phase of the feedback field, spin maser with continuous spin oscillations lasting hours can be realized \cite{Yoshimi2002PLA}. In further, if the feedback field is modulated by a square wave, one can expect a pulse modulated spin maser, which actually makes a magnetic frequency comb. 

Inheriting the merits of precision atomic magnetometers, the magnetic frequency comb works in an ultralow frequency range (at tens of Hertz level) and reaches high relative spectral resolution, with a quality factor of Q $\sim 10^4$. Due to the efficient spin feedback field coupling, the number of comb lines can easily go up to 50 with fine tuned pulse feedback parameters. A simple theoretical model describing the spectral envelope of frequency comb distribution is presented and agrees well with the experimental results. 

\begin{figure*}[htbp]
\centering
\includegraphics[width=1\textwidth,trim=0 0 0 0, clip]{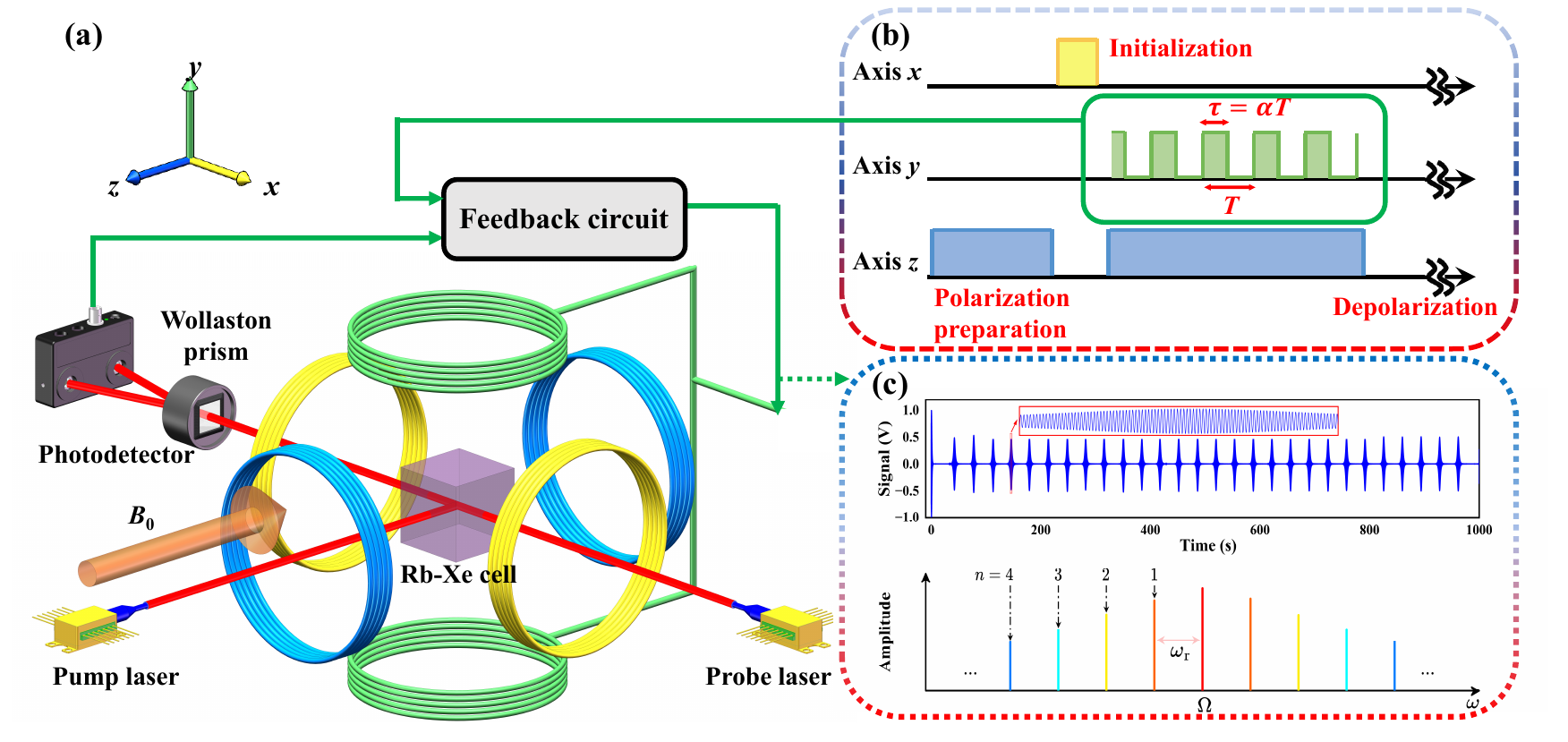}
\caption{\textbf{Experimental apparatus and time sequence protocol.} \textbf{(a)} The vapor cell in the center of magnetic shieldings contains a droplet of natural abundance Rb, 10 torr isotope enriched $^{129}$Xe and 50 torr of buffer gas N$_2$. The wavelengths of the pump and probe lasers are 795 nm and 780 nm, corresponding to the $D_1$ and $D_2$ lines of $^{87}$Rb. Specifically, the pump laser is locked at the transition frequency by a polarization spectroscopy. \textbf{(b)} In one experiment realization, we set the duration of polarization and depolarization to 60 s. In the initialization stage, $I_x$ is set to 1000 $\mu$A (current-to-magnetic coefficient $C_\mathrm{B/I}\sim$ 1.55 mG/mA) for 300 ms. The duty cycle $\alpha$ is set to 50$\%$.}
\label{fig1}
\end{figure*}

\textit{Experimental setup}--The experimental setup is depicted in FIG.\ \ref{fig1}. A 10$\times$10$\times$10 mm$^3$ cubic vapor cell consisting of a droplet of natural abundance rubidium, 10 torr isotope enriched $^{129}$Xe and 50 torr buffer gas $N_2$ is located at the center of a 4-layer magnetic shieldings. A dc magnetic field $B_0$ applied along the $z$-axis provides the quantum axis for spin precession. At a temperature of 120 $^\circ$C, the atomic spin ensemble in vapor states is optically pumped by a circularly polarized laser resonant with the Rb atomic $D_1$ transition line. The resultant $^{87}$Rb spin polarization is subsequently transferred to Xe via spin-exchange collisions, causing the Xe nuclear spins polarized. In the present experimental conditions, the Xenon spins have a coherence time of $T_2$$\sim5.64$ seconds in a typical free induction decay measurement.

A linearly polarized laser detuned from Rb $D_2$ transition line is used to probe the atomic spin precession signals, with the optical polarimetry configuration \cite{RMP2002Budker}. The above setup conforms a standard alkali-metal noble gas atomic comagnetometer. The major part to construct a pulse modulated spin maser is a specifically designed feedback circuit with an analog switch integrated on it. The circuit outputs a current driving the magnetic coil in $y$ direction. By tuning the amplitude and phase shift of the current, the driving field can excite the spin precession coherently and realize a stable spin maser. The pulse modulation function is realized by sending a programmable square wave signal to the analog switch.

The time sequence of a typical experiment cycle is illustrated in FIG.\ \ref{fig1}. The sequence starts with a polarization preparation by turning on the pump laser and $\bf B_0$ field in $z$ direction. Subsequently, a magnetic pulse $B_x$ in $x$ direction rotates the spin polarization $\bf M$ from $z$ direction to the $xy$ plane, then the nonzero torque imposed by $\bf B_0$$\sim$30 mG on $\bf M$ leads to the spin Larmor precession at frequency $\Omega=\gamma B_0$$\sim$35 Hz. Now the analog switch is turned on/off by a DAQ card (National Instrument USB-6351), imposing a pulse modulation on the feedback magnetic field $B_f$. The period $T$ and high-level duration $\tau$ of the modulation pulse is controlled by a LabVIEW program.

\textit{Experimental results}--FIG.\ \ref{fig1} shows the frequency comb spectra of the pulse modulated spin maser. Centered at the Larmor frequency, the highest peak is defined as the 0th-order comb line. The $n$th order comb lines ($n$$\ge$1) are distributed symmetrically around the 0th-order comb line and their amplitudes decrease progressively. The gap between the $n$th-order comb line and the 0th-order comb line is $\Delta \omega = n\omega_r$, where the repetition rate $\omega_r=2\pi/T_\mathrm{rep}$ is determined by the pulse period. 

\begin{figure}[htbp]

\centering
\includegraphics[width=0.5\textwidth,trim=30 0 30 0, clip]{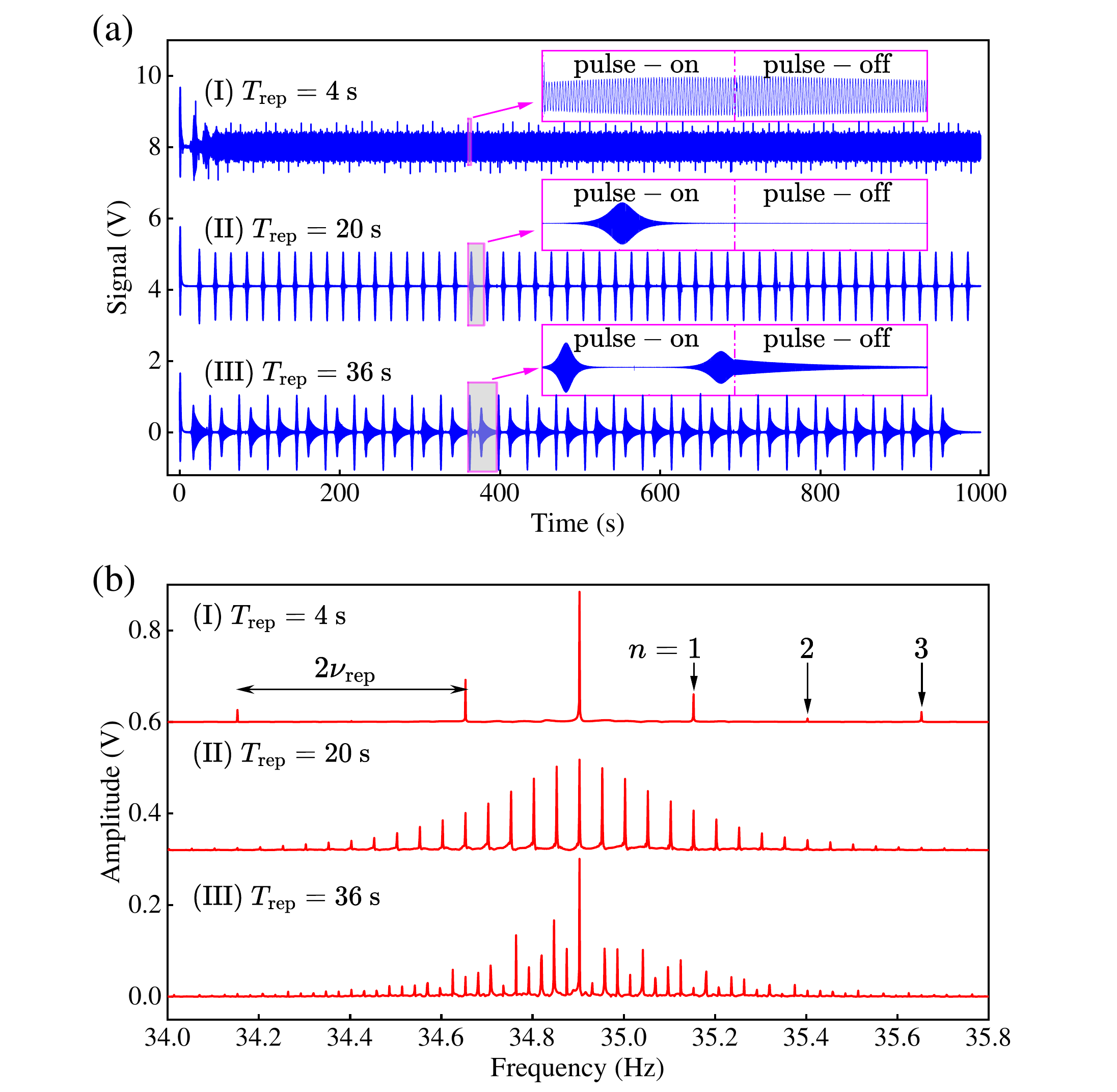}
\caption{\textbf{Characterization of Magnetic Frequency Combs (MFCs).}
(\textbf{a}) Representative time-domain signals of the MFCs at $T_{\mathrm{rep}} = 4~\mathrm{s}$, $20~\mathrm{s}$, and $36~\mathrm{s}$, corresponding to regimes \text{(I)}, \text{(II)} and \text{(III)}, respectively. 
(\textbf{b}) Frequency-domain spectra of the MFCs for the same pulse intervals shown in (\text{a}), illustrating the spectral evolution across the three characteristic regimes.}
\label{fig2}
\end{figure}

In experiments, it was found that while the creation of magnetic frequency comb relies on a critical feedback field strength, the comb spectral structure is generally insensitive to the strength $B_f$ and phase shift $\theta$ of the  feedback field once it emerges. The shape of the magnetic comb spectra is mainly affected by the pulse period $T_\mathrm{rep}$. We then analyze the MFC's spectral structure under different pulse period $T_\mathrm{rep}$. FIG. \ref{fig2} shows the frequency comb spectra with varying $T_\mathrm{rep}$ at a fixed strength of $B_f=0.024 B_0$$\sim$0.7 mG and $\theta=70^{\circ}$ (both are calibrated when the feedback circuit is in the open loop mode, i.e. its link to $y$-coil is off). Based on the amplitude distribution profiles of the comb sidebands, the spectra can be categorized into three distinct regimes: \textbf{Regime (I)}: In the region of small $T_\mathrm{rep}$, the majority of the energy is concentrated at the central frequency. The number of comb teeth is limited, and their amplitude distribution follows a Sinc function--like profile. \textbf{Regime (II)}: As $T_\mathrm{rep}$ gradually increases, the comb-tooth structure becomes increasingly pronounced, and the number of teeth grows significantly. In this regime, the amplitude distribution of the comb teeth resembles a Gaussian function--like profile. \textbf{Regime (III)}: When $T_\mathrm{rep}$ increases beyond a certain point, the frequency comb spectra become disorder with new lines emerging randomly.

To find the evolution law of the observed pulse spin maser dynamics and frequency spectra, we present a simple theoretical model by considering the fundamental principle of the feedback spin maser. The feedback mode of spin maser plays a key role in the formation of the magnetic frequency comb structures in FIG.\ \ref{fig2}. A spin maser starts with the partial rephasing of dephasing spin oscillations, like the spin echo with inhomogeneous dephasing and radio-frequency excitation rephasing \cite{Hahn1950PR,Meiboom1958RSI}. Here the feedback magnetic field plays the role of the radio-frequency rephasing field. In a continuous spin maser, the feedback field is continuous, which thus replenishes the transverse coherence of spin magnetization and creates multiple spin echoes. When the coherent driving effect of the feedback field balances the intrinsic decoherence losses of spin magnetization, the spin precession will evolve into a steady state of self-sustaining oscillations without any decay \cite{Yoshimi2002PLA}. Different from the continuous spin maser, pulse spin maser imposes a temporal gate that truncates the spin echoes and feeds them back to the spin oscillations. The way of this gate truncations determines how the spin oscillations behave ultimately and thus shapes the spectral structure of MFC. The truncation pulse can be written as 
\begin{equation}
\begin{split}
p(t)=\frac{A}{\pi} \sum_{n=-\infty}^{\infty} \frac{\sin(\alpha n \pi)}{n}\cdot e^{i n \omega_r t},
\end{split}
\label{eq1}
\end{equation}
\\where $\alpha$ represents the duty cycle, $A$ represents the amplitude, and $\omega_r$ is the repetition rate ($\omega_r=\frac{2\pi}{T_\mathrm{rep}}$). 
Suppose the original spin oscillation is $\chi(t)$. Here, $\chi(t)$ can be considered as a combination of the precession function and the envelope function $\chi(t)={\xi(t)}\cdot e^{i\Omega t}$, where $\Omega=\gamma B_0$ represents the Larmor precession frequency, and $\xi(t)$ is the time domain envelope of magnetization signal $\chi(t)$. Then we define $\Omega_n^\pm = \Omega \pm n \omega_r$ for the sake of simplicity. Consequently, we get the feedback field as
\begin{equation}
\begin{aligned}
B_f (t)
& =\chi(t) \cdot p(t) \\ 
&=\xi(t) \cdot \alpha A \cdot [e^{i \Omega t}+\sum_{n=1}^{\infty}(e^{i\Omega_n^+ t}+e^{i\Omega_n^- t})\cdot \operatorname{sinc}(\alpha n \pi)].
\end{aligned}
\label{eq2}
\end{equation}

\begin{figure*}[hbtp]
\centering
\includegraphics[width=1\textwidth,trim=30 0 30 0, clip]{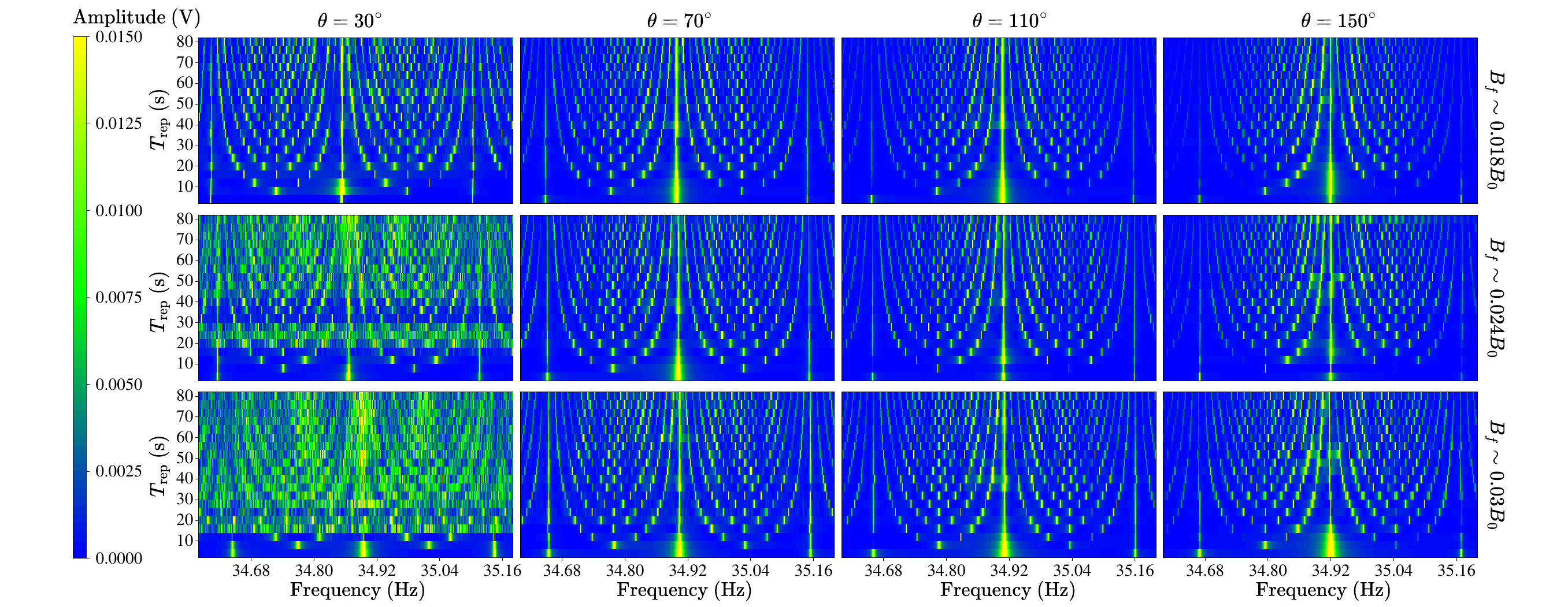}
\caption{\textbf{Intensity heatmap of the magnetic frequency comb for various pulse durations under different strengths and phases of feedback field.} The magnetic frequency comb reaches its best performance at empirical ranges of the three key control parameters: $B_f$ $\sim$ (0.01,0.1) $B_0$, $\theta$ $\sim90\pm20^\circ$ and $T_\mathrm{rep}$ $\sim$ 3-6 $T_2$, where the frequency comb has largest number of teeth and equal frequency spacing. The comb spectra at phase shift $\theta$= 30$^\circ$ and 150$^\circ$ (significant phase mismatch) show distortion partly or mostly, indicating the presence of nonlinear dynamics in the pulse modulated spin maser.}
\label{fig3}
\end{figure*}

With Fourier transformation, the frequency spectra of $B_f (t)$ is expressed as
\begin{equation}
\begin{aligned}
S_f(\omega)
&=\mathcal{F}\{B_f (t)\} \\
&=\mathcal{F}\{\xi(t)\} \cdot 2\pi \alpha A \cdot \{\delta(\omega-\Omega) \\
&\quad +\sum_{n=1}^{\infty} \operatorname{sinc}(\alpha n \pi) \cdot [\delta(\omega-\Omega_n^+)+\delta(\omega-\Omega_n^-)] \} \\
&\sim \delta(\omega-\Omega)+\sum_{n=1}^{\infty} \operatorname{sinc}(\alpha n \pi)\cdot [\delta(\omega-\Omega_n^+)+\delta(\omega-\Omega_n^-)],
\end{aligned}
\label{eq3}
\end{equation}
\\where $\mathcal{F}$ represents the Fourier transform, and $\delta$ is the Dirac function.

The equation \ref{eq3} explains the frequency comb spectra structure well in details. The position of the $n$th-order comb line is given by $\omega=\Omega_n^\pm$. The term $\sum_{n=1}^{\infty} \operatorname{sinc}(\alpha n \pi)$ indicates that the intensities of comb lines follow a distribution of the sinc function. Specifically, in our experiments, with $\alpha=0.5$, each value of $n$ meeting the equation $n=2k$ ($k$ is a positive integer) will make $\operatorname{sinc}(\alpha n \pi)=0$, resulting in the elimination of the even-order comb line in FIG.\ \ref{fig2}b (I).
  
When $T_\mathrm{rep}$ is shorter than the duration of a single spin echo, which is about the spin decoherence time $T_2$, the gate truncation of spin echo is frequent. A single spin echo does not finish its entire period before another echo comes in. In this case, the pulsed spin maser works in a quasi-continuous mode, which is weakly modulated by a square wave. Weak modulation means fewer sidebands, therefore the spectral envelope of the MFC exhibits a Sinc function-like distribution with only a few nonzero order comb lines as shown in FIG.\ \ref{fig2}b ($T_{\mathrm{rep}}=4$ s). 

When $T_\mathrm{rep}$ becomes comparable to the duration of a single spin echo, a complete single spin echo is realized within one pulse period in FIG.\ \ref{fig2}a ($T_{\mathrm{rep}}=20$ s). Given a strong enough feedback field, the spin echo can repeat itself cycle by cycle with a shifted phase each cycle determined by the phase shift $\theta$. In the frequency domain, this chain of spin echoes can be interpreted by the convolution of an infinite number of Sinc function distributions, which converges to a Gaussian function distribution (according to the central limit theorem) in the comb spectra. This agrees well with the experimental results for $T_{\mathrm{rep}}=20$ s in FIG.\ \ref{fig2}b.

If the pulse duration increases further until it surpasses the duration of a single spin echo, a second spin echo will enter the feedback pulse. This second spin echo will change the phase of the spin oscillations in an abrupt way, thus breaking the phase synchronization. Therefore, the overall chain of spin echoes will have random phases. In frequency domain, the MFC spectra become disordered  in FIG.\ \ref{fig2}b (III). According to above analysis and experimental results, the boundary between regimes (II) and (III) can be set by the emergence of the second-echo in the pulse spin maser dynamics.

To have a full view on the property of the magnetic frequency comb, we changed the three important parameters: $B_f$, $\theta$ and $T_\mathrm{rep}$ point by point in a broad parameter space and obtained the corresponding spectra. The heatmap of the spectral amplitude of MFC versus the three parameters are plotted in FIG.\ \ref{fig3}. In principle, the stronger the feedback field, the easier the frequency comb produced, therefore one can see the number of effective comb lines increases as $B_f$ increases. The minimum feedback field strength for the notable production of magnetic frequency comb is $B_f=0.01 B_0$$\sim 0.3$ mG, about 7.5 times that for the continuous spin maser, which is reasonable considering the driving energy loss and other factors such as the impedance mismatch in pulse feedback mode \cite{Li2023PRAppl}.

The phase shift $\theta$ gives the relative phase between the feedback field and the spin-precession signal and thus determines whether the spin echoes interfere with each other constructively or destructively. As we use a quadrature feedback configuration (optical detection in the $x$ direction and feedback in the $y$ direction), the feedback field drives the spin Larmor precession in phase at $\theta=$ 90$^\circ$.
 As a result, the formation of stable spin echo chain shall be possible only within a limited narrow phase shift window centering at $\theta=$ 90$^\circ$. However, due to the slightly nonlinear spin maser dynamics, there are also spin maser effects in a finite phase shift window, and the higher the feedback field strength, the broader the phase shift window \cite{Li2023PRAppl,Feng2025PRA}. 

On the other hand, however, the strength and phase shift of feedback field can not change too much because the nonlinear spin maser dynamics may happen. As shown in the bottom left area of the Fig. \ref{fig3}, there is notable bright region in the magnetic frequency comb spectra at $\theta=$ 30$^\circ$. It is a visual effect due to the spreading and crowding of the maser oscillation modes over a broad frequency range. The frequencies of these oscillation modes are randomly distributed. This phenonmena is due to the presence of nonlinear dynamics at large phase mismatch in the strong feedback spin masers \cite{Feng2025PRA}. In further, this nonlinear effects is asymmetric about $\theta=$ 90$^\circ$, as it does not happen at $\theta=$ 150$^\circ$. The asymmetry of nonlinear effects is probably linked to the polarity dependence of the artificial feedback spin masers, we have proved this dependence by switching the two ends of the feedback circuit output in the initial operation of pulse spin maser \cite{Jiang2021SA}. 

To conclude, in order to make the magnetic frequency comb working effectively, one need to tune the feedback field to a modest strength $B_f$ $\sim$ (0.01,0.1) $B_0$, the phase shift within a range of $\theta$ $\sim90\pm20^\circ$ in the quadrature feedback configuration and the pulse duration to several times the spin coherence time $T_\mathrm{rep}$ $\sim$ 3-6 $T_2$. In addition, one needs to keep in mind that these parameters need to be tuned cooperatively to avoid the nonlinear effects in spin masers. In this parameter space, the magnetic frequency comb using pulse spin maser can have the largest number of usable comb lines ($n$th order comb lines with amplitude over $5\%$ that of the zeroth order comb line) and linear response. The number of effective comb lines can go up to 50 and the quality factor reaches $10^4$, indicating the high performance of realized magnetic frequency comb. The linewidth of the comb lines can reach sub-millihertz and obtain a frequency resolution down to tens of nanohertz (nHz). The magnetic frequency comb can be applied in areas needing broadband ultrasensitive magnetic sensors, such as the search for spin-dependent exotic physics and ultraweak biomagnetic field detections \cite{Newphys2018RMP,Terrano2022QST,MEG2018Nature}.

\textit{Conclusion}--In summary, we realized an ultralow frequency magnetic comb using the pulse modulated spin maser in a gaseous Rb-Xe hybrid spin ensemble. With long nuclear spin coherence time and efficient spin-feedback field coupling, the frequency comb exhibits ultra high frequency resolution and broad spectral range. In principle, the demonstrated pulse modulated magnetic frequency comb can be extended to pure alkali-metal spin masers such as Rb or Cs spin masers, which can work at higher carrier frequency (tens of kilohertz or higher) and find applications in radio-frequency or microwave precision magnetic measurements. 

The authors would like to appreciate the financial support by the Chinese Academy of Sciences under grant number E209YC1101.

\end{document}